\documentstyle[aps,prb]{revtex} 
\begin{document}
\title{Small-World Rouse Networks as models of cross-linked polymers}
\author{S. Jespersen$^{a,b}$}
  \author{I. M. Sokolov$^b$ and A. Blumen$^{b}$}
\address{$^a$Institute of Physics and
  Astronomy\\ 
University of Aarhus, DK-8000 {\AA}rhus C}
\address{$^b$Theoretische
Polymerphysik\\ Universit\"at Freiburg, D-79104 Freiburg i.Br.,
Germany\\ }
\date{\today}   
\maketitle

\begin{abstract}
\large{{\it We use the recently introduced small-world networks (SWN)
  to model cross-linked polymers, as an extension of the linear
  Rouse-chain. We study the SWN-dynamics under the influence of
  external forces. Our focus is on the structurally and thermally
  averaged SWN stretching, which we determine both numerically and
  analytically using a psudo-gap ansatz for the SWN-density of
  states. The SWN stretching is related to the probability of a
  random-walker to return to its origin on the SWN. We compare our
  results to the corresponding ones for Cayley trees.
}}
\end{abstract}
\newpage
\large{
\section{Introduction}
Topological properties of polymers can dramatically effect their
dynamical properties, such as their collapse in bad solvents and their
response to external forces
\cite{sommer,schiessel1,schiessel2,parbati,kantor}. Such forces can be
applied microscopically, either by having charged polymers
(polyelectrolytes, polyampholytes) in electrical fields, or via
optical tweezers or magnetic beads\cite{quake,perkins,wirtz}. In this
communication we study the stretching of cross-linked objects, whose
backbones are regular lattices (we will consider for simplicity a
ring), a few elements of the backbone being chemically
connected to each other via cross-links. An experimental realization
may be a very dilute solution of linear chains which are then
cross-linked by irradiation\cite{klaumunzer} or chemically. Our model is a
realization of the so-called small-world
networks\cite{watts,sj,monasson,newman1,moukarzel2,newman2} (SWN), and
the disorder (the cross-links) is in statistical terms quenched.
Considering all bonds to be equal we study the dynamics in the framework of
the Rouse model\cite{rouse,doi}, in which the monomers are connected
by harmonic springs; we term this structure the small-world Rouse
network (SWRN). Our SWRN is built out of an N-monomer ring with
superimposed fixed links between randomly chosen monomers; no
additional links are generated or broken in external fields. The SWRN
is a new intermediate between linear chains and networks. Distinct
from Cayley-trees, which model hyperbranched polymers without loops,
loops are a fundamental ingredient here. As such the SWRN is
interesting in its own right as a study of the interplay between
dynamics and topology, and it belongs to the class of generalized
Gaussian structures\cite{sommer,schiessel1,schiessel2}.

\section{Dynamics of small-world Rouse networks}
The construction of the SWN which we consider here (see also
\cite{sj}), is slightly different from the original one by Watts and
Strogatz\cite{watts}, but preserves its main
SWN-characteristics. Starting from a ring of $N$ monomers,  we
cross-link with probability $p$ each monomer randomly to any of the
monomers of the network. Such cross-links thus connect monomers far
apart along the chemical backbone (the ring), rendering them close in
Eucledian space. The SWRN-monomers are, in accordance with the Rouse
model\cite{rouse,doi}, connected by harmonic springs of strength
$k$.   The position  ${\mathbf R}_n(t)$ of the $n$th bead under the
action of an external force ${\mathbf F}_n(t)$ and in the presence of
thermal noise ${\mbox{\boldmath $\eta$}}_n(t)$ is governed by the
Langevin equation
\begin{equation}
\label{langevin1}
\gamma \frac{d{\mathbf R}_n(t)}{dt}=k\sum_{j=1}^N A_{nj}{\mathbf
  R}_j(t)+{\mathbf F}_n(t)+{\mbox{\boldmath $\eta$}}_n(t).
\end{equation}
Here $\gamma$ is the coefficient of friction, and the matrix $A_{ij}$
is the connectivity matrix of the network. It is defined as follows:
Every connection between site $i$ and site $j$ contributes $-1$ to
$A_{ij}$, while $A_{ii}$ and $A_{jj}$ are determined from the
condition that $\sum_j A_{ij}=\sum_i A_{ij}=0$.  Monasson
\cite{monasson} recently published a detailed study of the spectrum of
the  small-world connectivity-matrix $A$. Among his findings was the
existence of a ``pseudo-gap'' in the SWN density of states $\rho(E)$
which has the form
\begin{equation}
\label{pseudo}
\rho(E)\sim E^{-1/2} \exp\left(-\frac{C}{\sqrt{E}}\right),\hspace{1cm}
E\rightarrow 0
\end{equation}
This behavior could not be confirmed numerically through direct
diagonlization, and appeared in the data as a real gap. There are, as
we shall see, other ways of probing this behavior numerically, and our
results support that $\rho(E)$ behaves as Eq. (\ref{pseudo}).  

Writing
\begin{equation}
\label{notation}
{\mathbf R}(t)\equiv \left({\mathbf R}_1(t),{\mathbf
  R}_2(t),\ldots,{\mathbf R}_N(t)\right)^T
\end{equation}
and similary for the other quantities in Eq. (\ref{langevin1}), we can
rewrite the equation of motion in a more condensed form as
\begin{equation}
\label{langevin2}
\frac{d{\mathbf R}(t)}{dt}=\sigma A{\mathbf R}(t)+\frac{{\mathbf
    F}(t)}{\gamma}+{\mathbf w}(t).
\end{equation}
Here we have introduced $\sigma\equiv k/\gamma$ and ${\mathbf w}\equiv
{\mbox{\boldmath$\eta$}}/\gamma$. In the case of a spatially constant
external force, the solution to Eq. (\ref{langevin2}) is obtained as 
\begin{equation}
\label{solution}
{\mathbf R}(t)=\int_{-\infty}^t ds\, e^{-\sigma A (t-s)}\left(
\frac{{\mathbf F}(s)}{\gamma}+{\mathbf w}(s)\right)
\end{equation}
We now specialize to the following situation: the force is switched on
at time $t=0$ and pulls only the $m$th bead in the $y$-direction, i.e.
${\mathbf F}_i(t)=\theta(t)\delta_{i,m}F\,\hat{{\mathbf y}}$. Here
$\theta(t)$ is the Heaviside step-function, $\delta_{i,j}$ Kronecker's
delta and $\hat{{\mathbf y}}$ is a unit-vector pointing in the
$y$-direction. We focus on the displacement of the $m$th bead along
the $y$-axis, and average over thermal noise, using $\langle {\mathbf
w}(t)\rangle =0$. Finally we perform a structural average over $m$ and
end up with (see e.g. \cite{schiessel1,schiessel2,parbati} and
references therin for details)  
\begin{equation}
\label{extension}
Y(t)\equiv \frac{1}{N}\sum_m\langle{\mathbf R}_{m,y}(t)\rangle
=\frac{F}{N\gamma}\int_0^t ds\, \sum_i e^{-\sigma \lambda_i
s}=\frac{Ft}{N\gamma}+\frac{F}{N\gamma\sigma}\sum_{i=2}^N
\frac{1-e^{-\sigma \lambda_i t}}{\lambda_i}.
\end{equation}
In this equation ${\mathbf R}_{m,y}$ is the $y$ component of ${\mathbf
R}_m$ and $\lambda_i$ with $i=1\ldots N$ are the eigenvalues of the
connectivity matrix $A$. The last expression follows from the fact
that for a connected structure, only one eigenvalue vanishes (say,
$\lambda_1$). At times $t$ much smaller than the time scale set by the
largest eigenvalue $\lambda_{max}$, i.e. when $\sigma\lambda_{max}
t\ll 1$, $Y(t)$ increases linearly in time: $Y(t)\sim Ft/\gamma$. That
is, only the monomer being pulled moves with a constant speed, not yet
feeling the influence of the other monomers. Likewise, at late times
$\sigma\lambda_{min} t\gg 1$, where $\lambda_{min}$ is the lowest
non-vanishing eigenvalue, the entire polymer is being pulled with a
constant speed, $Y(t)\sim Ft/(N\gamma)$. These observations are
independent of the specific structure being pulled, and only in the
intermediary regime $\lambda_{max}^{-1}\ll \sigma
t\ll\lambda_{min}^{-1}$ does the particular topology of the polymer
affect the dynamics, namely through the spectrum of the connectivity matrix. 

The numerical computation of the quantity $Y(t)$ above proceeds as
follows. From a specific realization of an $N=1000$ small-world
network, we construct the corresponding connectivity matrix. Then we
find the $N$ eigenvalues using standard routines, and implement
Eq. (\ref{extension}). To get an idea of the importance of sample to
sample fluctuations, we consider first $10$ different realisations of
the SWRN for $p=0.05$. Plotted is in Fig. \ref{samples} on double
logarithmic scales $Y$ as a function of $t$, where here and in the
following we use the dimensionless variables $Y^*(t)\equiv
\sigma\gamma Y(t)/F$ and $t^*\equiv\sigma t$. In Fig. \ref{samples} we
display the envelope of all $10$ realizations, i.e. the two curves are
the extremal two ``worst'' cases. 
We see that the difference in the results is quite small (and appears,
as it should, only for intermediary times), and therefore we regard
results from any specific realization as being typical. In
Fig. \ref{comp1} we analyze the dependence of $Y(t)$ on cross-linking,
by varying $p$ from $p=0$ (standard Rouse-model of the ring) to
$p=0.01$ and $p=0.05$. 
We note first that the differences are now considerably larger than in
Fig. \ref{samples}. Second, for $p=0$ i.e. for the Rouse chain, we
have the standard picture: a subdiffusive $\sqrt t$ behavior at
intermediary times is followed by a diffusive $t$ behavior at longer
times. At very early times we also have a linear behavior, albeit not
visible in the range of the figure. The initial and final dynamics are
in accordance with the explanation given above\cite{parbati}.
The intermediate behavior reflects the structure of the spectrum of
$A$, and is also well  understood in the Rouse case; it is a result of
the rather slow propagation of disturbances through the chain (here
the ring). 

We infer from the other curves in Fig. \ref{comp1}, that even very
small but nonvanishing $p$ affect the intermediate behavior of $Y(t)$
quite strongly. For increasing $p$ the curves bend downwards from the
$p=0$ case, 
mirroring the increased stiffness of the polymer due to the additional
links. Thus a ring with cross-links can be easily distinguished
experimentally (say through NMR or electronic energy transfer) from
one without cross-links, whose $Y(t)$ dynamics under $\mathbf{F}$ is
never slower than $\sqrt{t}$. As expected, the very early and very
late behavior in all three cases coincide, being independent of the
specific structure under scrutiny.

In Fig.  \ref{p} we plot on logarithmic scales $Y^*(t)$ as a function
of time $t^*$ for several values of $p$. Increasing $p$ increases the
stiffness of the polymer, and this is reflected in the intermediary
regime, which becomes almost flat for large $p$. Moreover the long
time behavior of the polymer is reached much earlier for polymers with
a higher number of cross-links. In line with the discussion above of the
range of the intermediary regime ($\lambda_{max}^{-1}\ll \sigma
t\ll\lambda_{min}^{-1}$), this feature means that the lowest
non-vanishing eigenvalue $\lambda_{min}$ gets quite large, and it is
hence related to the appearance of a (pseudo) gap in the spectrum of $A$.}  \large{
\section{Theoretical Analysis}
As indicated earlier, the initial as well as the asymptotic behavior of
$Y(t)$ are well understood. Thus we will concentrate here on the
richer and much more complex intermediate behavior.  We shall rewrite
Eq.~(\ref{extension}) using a continuous picture, based on the density
of states 
$\rho(\lambda)=\lim_{N\rightarrow\infty}(1/N)\sum_{i}
\delta(\lambda-\lambda_i)$, but continue to separate out the vanishing
eigenvalue from the rest. Hence, with $\epsilon$ very small,
$\epsilon\rightarrow 0^+$:   
\begin{equation}
\label{ext}
Y(t)=\frac{F}{N\gamma}\int_0^tds\, \sum_i e^{-\sigma \lambda_i s}
= \frac{Ft}{N\gamma}+\frac{F}{\gamma} \int_0^tds \,
\int_\epsilon^{\infty} d\lambda \,\rho(\lambda)e^{-\sigma\lambda s},
\end{equation}
an expression which is {\it a fortiori} correct in the presence of a
gap, where one can take $0<\epsilon\leq\lambda_{min}$. It will be convenient also to consider the {\em stretching} (relative
motion) $\Delta(t)$ separately: 
\begin{equation}
\label{asympext}
\Delta(t)\equiv Y(t)-\frac{Ft}{N\gamma}=\frac{F}{\gamma} \int_0^tds \,
\int_{0^+}^{\infty} d\lambda \,\rho(\lambda)e^{-\sigma\lambda s}.
\end{equation}
We remark that the inner integral in Eq. (\ref{asympext}) is related
to the probability for a random walker to be present at the original
site\cite{sj,bouchaud}. We therefore first analyse the  behavior of
this quantity:
\begin{equation}
\label{presence}
P_0(t)\equiv\int_0^\infty d\lambda\,\rho(\lambda)e^{-\lambda t}
\end{equation}
The asymptotic temporal behavior is accessed through the behavior of
$\rho(\lambda)$ at small $\lambda$. Inserting the expression of
Monasson\cite{monasson} Eq. (\ref{pseudo}) into Eq.~(\ref{presence})
we obtain:
\begin{equation}
\label{asymp1}
P_0(t)\sim\int_0^\infty d\lambda\,\lambda^{-1/2}
\exp\left(-\frac{C}{\sqrt{\lambda}}-\lambda
t\right)=-t^{-2/3}\frac{d}{dC}\int_0^\infty
dy\,\exp\left(-t^{1/3}(\frac{C}{\sqrt{y}}-y)\right) 
\end{equation}
The asymptotic behavior of the integral follows readily from a
saddle-point procedure \cite{asympt}, so that we end up with 
\begin{equation}
\label{asymp2}
P_0(t)\sim t^{-1/2}\exp(-C't^{1/3}),  
\end{equation}
whith $C'=3\left(C/2\right)^{2/3}$.This is by itself a quite interesting and novel result, and it
compares favourably to our numerical simulations\cite{sj} for
$P_0(t)$. The result is close in form to that for Cayley
trees\cite{cassi1,cassi2}, where $P_0(t)\sim
t^{-3/2}\exp(-ct)$. Inserting Eq. (\ref{asymp2}) in
Eq. (\ref{asympext}) and reintroducing $\sigma$, we get  
\begin{equation}
\label{predict}
\Delta(t) 
\sim
\frac{3F}{C'\gamma\sigma}\left(\frac{1}{2}\sqrt{\frac{\pi}{C'}}-(\sigma
t)^{1/6}e^{-C'(\sigma t)^{1/3}}\right).
\end{equation}
Notice that for very large $t$ the stretching $\Delta(t)$ of the SWRN
tends to a constant $\Delta_\infty$. In the units of our figure this
constant depends 
mainly on $C'$ (since $F$, $\gamma$ and $\sigma$ drop out);
theoretically one may obtain $C'$ and also $C$ out of $\Delta_\infty$. 

In Fig. \ref{theory} we plot the dimensionless stretching
$\Delta^*(t)\equiv Y^*(t)-\sigma t/N$ for $p=0.05$ and compare it to
the analytical form Eq.~(\ref{predict}). We do this by  fitting
$a-bt^{1/6}\exp(-ct^{1/3})$ to the data, and as can be inferred from
Fig.~\ref{theory}, the agreement is very convincing. From the
least-squares fit we obtain $a=5.09$, $b=7.67$ and $c=0.54$. We remark
that the agreement at short times may be rendered even better by also
keeping the next term in the expansion of the integral of $P_0(t)$, a
term which is proportional to $t^{-1/6}\exp \left({-C't^{1/3}}
\right)$. 
Furthermore we remark that for Cayley-trees\cite{parbati} the
intermediate behavior of $Y(t)$ can also be determined in a similar
manner: The saddle-point approach yields to leading order:
\begin{equation}
\label{cayley}
Y(t)\sim \tilde{a}-\tilde{b}t^{-3/2}\exp(-\tilde{c}t),
\end{equation}
with $\tilde{a}$, $\tilde{b}$ and $\tilde{c}$ being constants. 
\section{Conclusions}
In this communication we have studied the behavior of a small-world
network model (the SWRN) of a linked ring-polymer, focusing on its
dynamics under external forces. Thus the motion of a monomer pulled by
such a force is vastly different, depending on whether the monomer
belongs to a SWRN or to a simple ring without cross links. This may
enable via NMR or electronic energy transfer\cite{blumen} to
distinguish clearly between cross-linked and non-cross-linked
polymers.  Our numerical results for the stretching of the SWRN under
external forces are in excellent agreement with our analytical
expressions, which used the pseudo-gap behavior of the SWN density of
states, as postulated in former work\cite{monasson}. As discussed in
the present communication, these results are directly connected to
expressions for the return to the origin of a random walker on the SWN. 

\begin{acknowledgements}
The support of the DFG, of the GIF through grant I0423-061.14, and of
  the Fonds der Chemischen Industrie are gratefully acknowledged.
\end{acknowledgements}

}

\newpage
{\large FIGURES}
\begin{figure}[h]
\unitlength=1cm
\begin{center}
\begin{picture}(6,5.8)
\put(-2.2,6.6){ \includegraphics{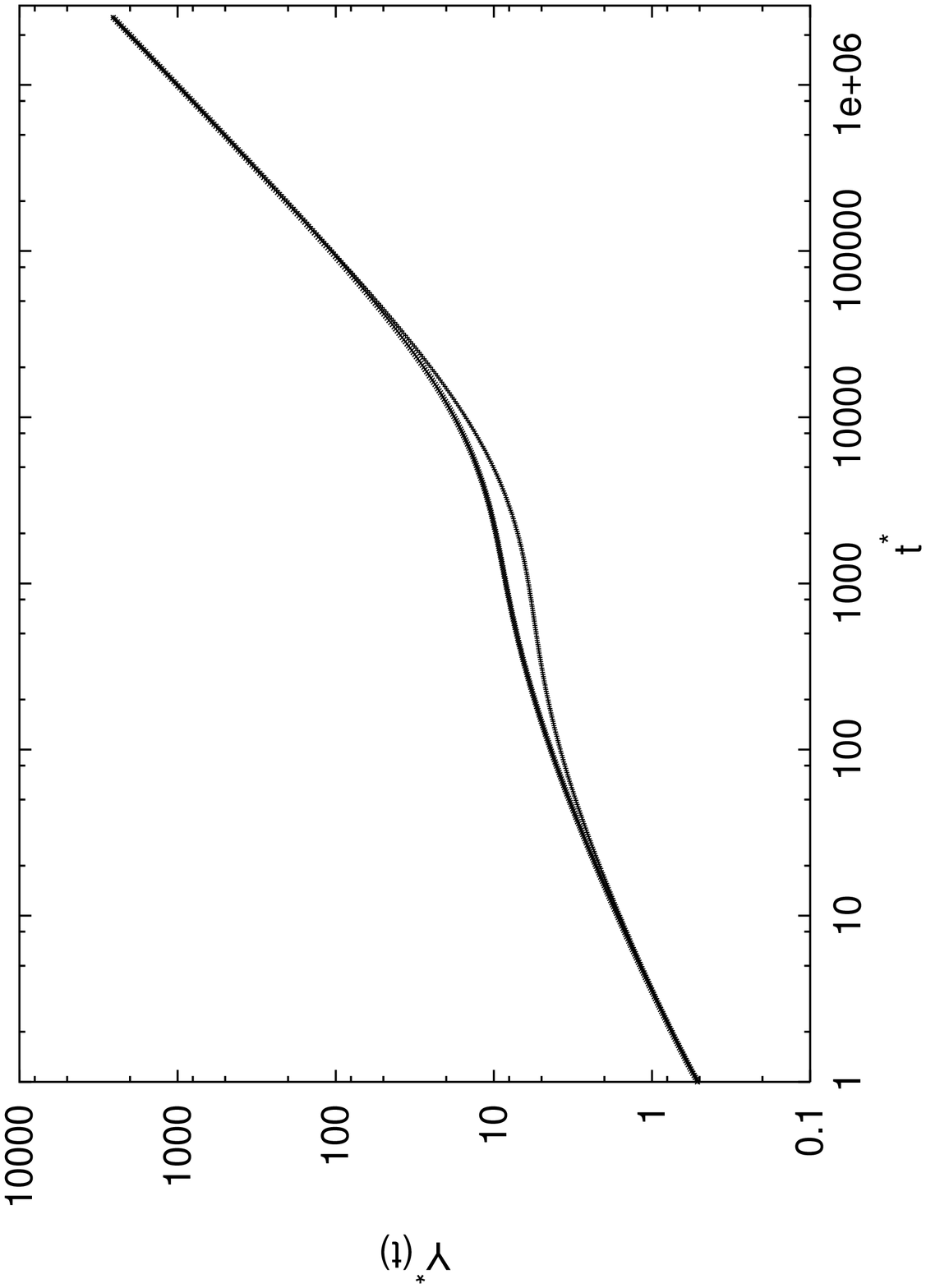}}
\end{picture}
\end{center}
\caption{}
\label{samples}
\end{figure} 

\begin{figure}[h]
\unitlength=1cm
\begin{center}
\begin{picture}(6,5.8)
\put(-2.2,6.6){ \includegraphics{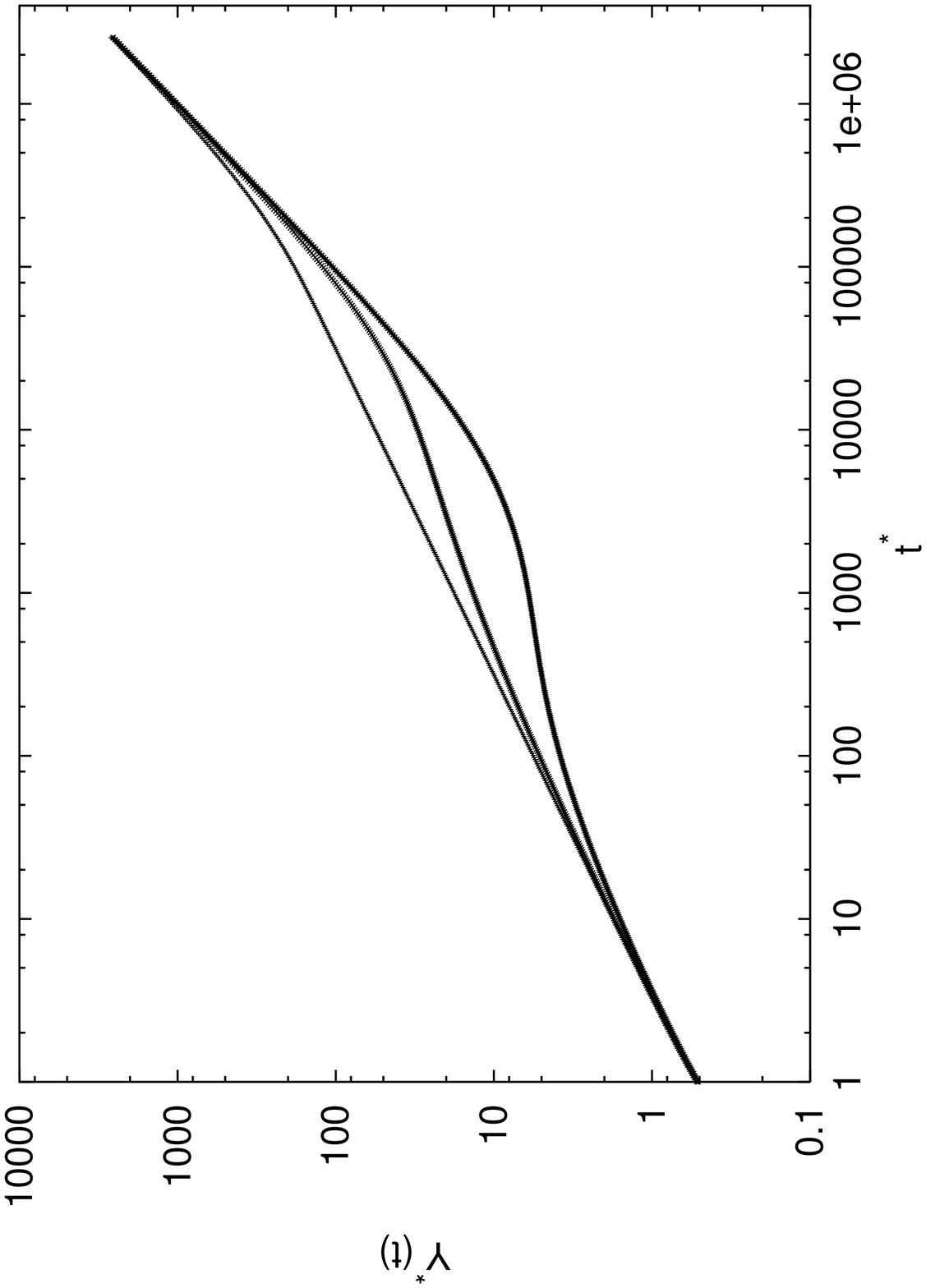}}
\end{picture}
\end{center}
\caption{}
\label{comp1}
\end{figure} 

\begin{figure}[h]
\unitlength=1cm
\begin{center}
\begin{picture}(6,5.8)
\put(-2.2,6.6){ \includegraphics{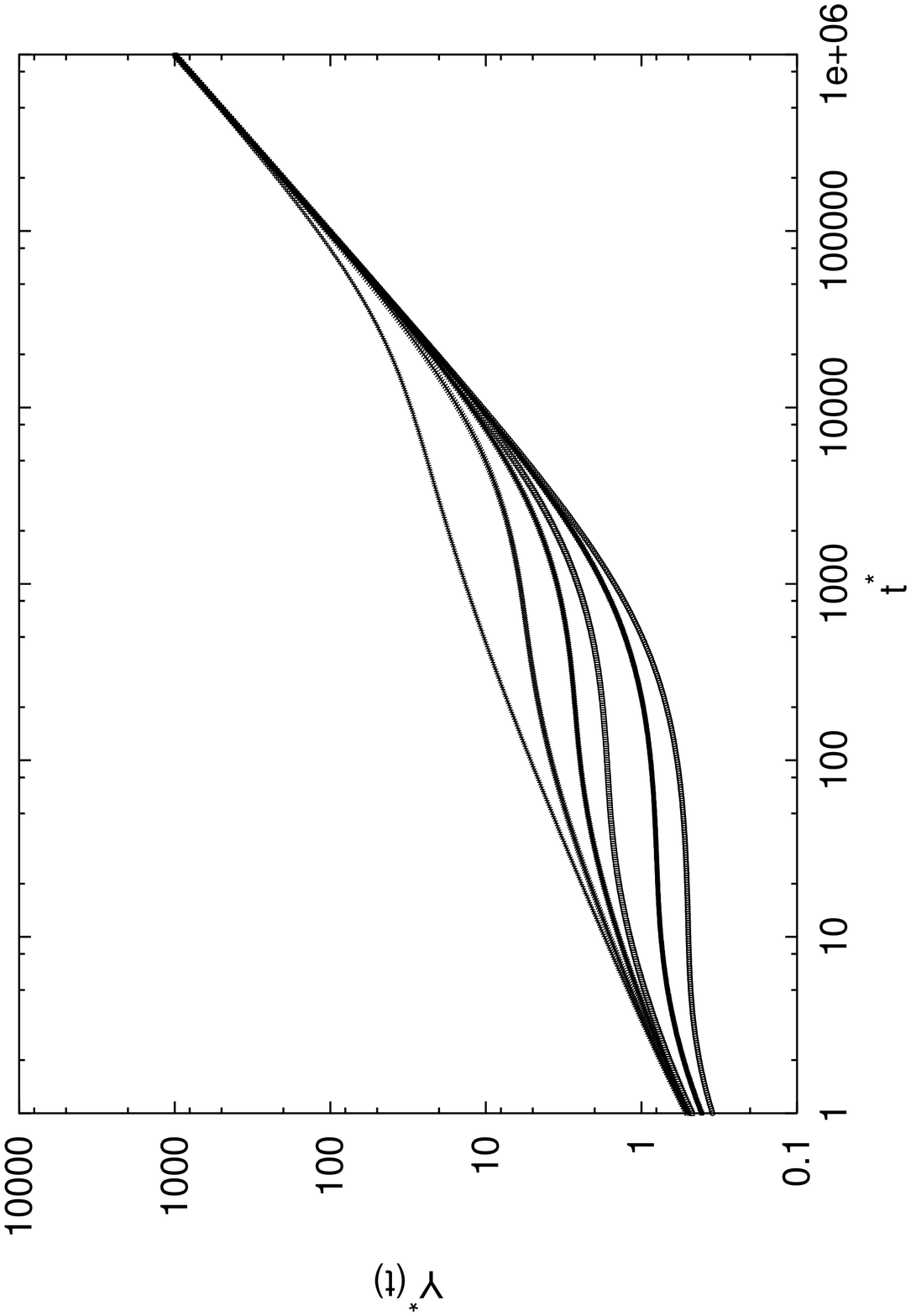}}
\end{picture}
\end{center}
\caption{}
\label{p}
\end{figure} 

\begin{figure}[h]
\unitlength=1cm
\begin{center}
\begin{picture}(6,5.8)
\put(-2.2,6.6){ \includegraphics{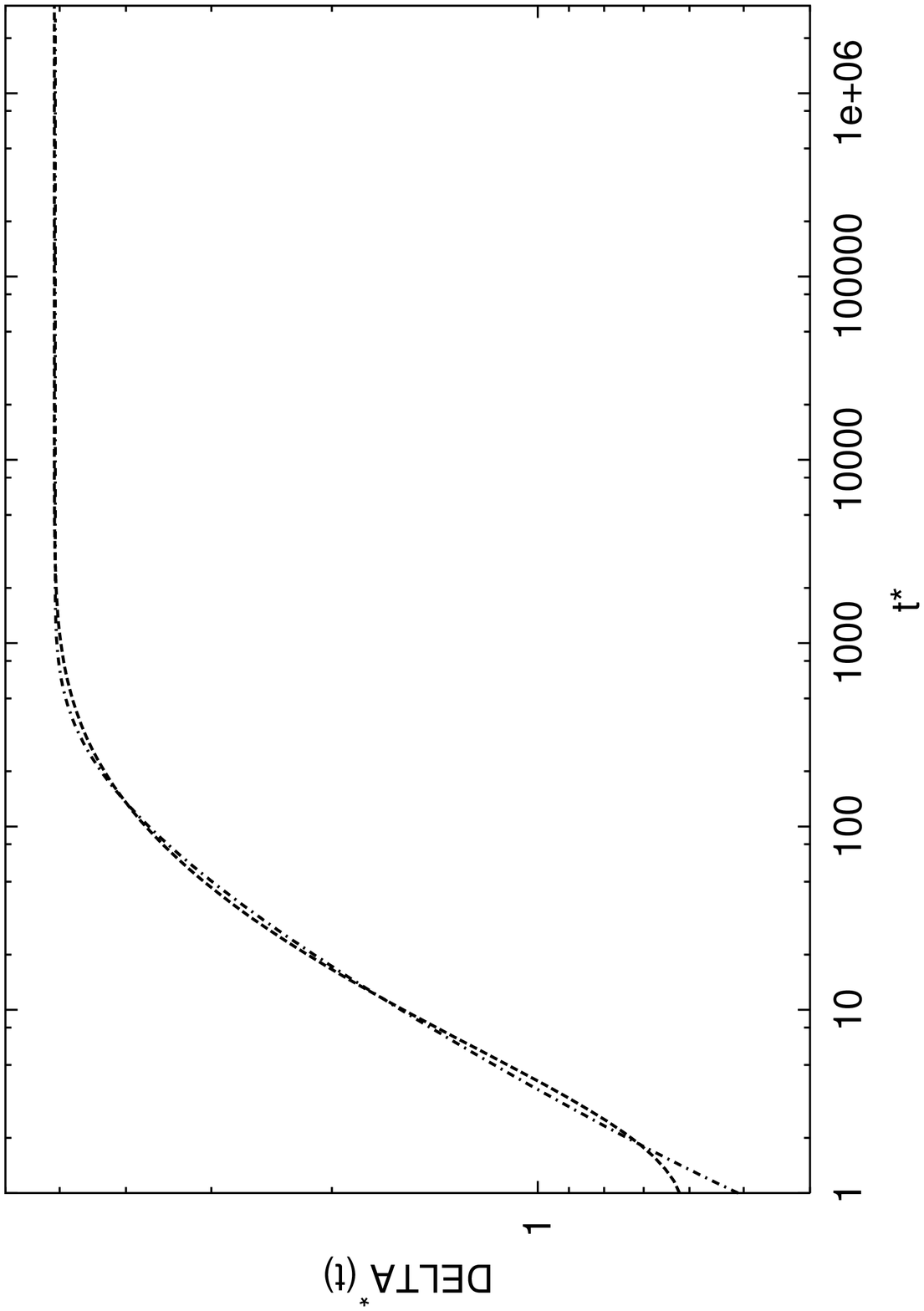}}
\end{picture}
\end{center}
\caption{}
\label{theory}
\end{figure}

\newpage
{\large CAPTIONS}\\

\noindent FIG. 1. Two different realizations give rise to similar
  behavior of $Y(t)$, here plotted on logarithmic-logarithmic scales
  for $p=0.05$.
\\

\noindent FIG. 2.  On double logarithmic scales we plot the position
  $Y(t)$ as a function of time. From upper to lower curve, $p=0$, $p=0.01$ and
  $p=0.05$.
\\

\noindent FIG. 3. On double logarithmic scales we plot the position $Y(t)$ as a
  function of time. From upper to lower curve, $p=0.01$, $p=0.05$,
  $p=0.1$, $p=0.2$, $p=0.5$ and $p=0.8$.
\\

\noindent FIG. 4. Comparison of the theoretical prediction (dashed) with
  the data (dash-dotted), for $p=0.05$.

\end{document}